\documentclass[twocolumn,showpacs,preprintnumbers,amsmath,amssymb, showkeys]{revtex4}

\usepackage{graphicx}
\usepackage{dcolumn}
\usepackage{bm}
\usepackage{epsf}

\begin{document}


\title{Tuning of tunneling current noise spectra singularities by localized states charging}

\author{V.\,N.\,Mantsevich}
 \altaffiliation{vmantsev@yahoo.com}
\author{N.\,S.\,Maslova}%
 \email{spm@spmlab.phys.msu.ru}
\affiliation{%
 Moscow State University, Department of  Physics,
119991 Moscow, Russia
}%

\date{\today }

\begin{abstract}
We report the results of theoretical investigations of tunneling
current noise spectra in a wide range of applied bias voltage.
Localized states of individual impurity atoms play an important role
in tunneling current noise formation. It was found that switching
"on" and "off" of Coulomb interaction of conduction electrons with
two charged localized states results in power law singularity of
low-frequency tunneling current noise spectrum ($1/f^{\alpha}$) and
also results on high frequency component of tunneling current
spectra (singular peaks appear).

\end{abstract}

\maketitle

\section{Introduction}

   In the present work we discuss one of the possible reasons for the
noise formation in the wide range of applied bias in the STM/STS
junctions. Up to now the physical nature and the microscopic origin
of tunneling current noise spectra formation in low and high
frequency region is in general unknown. Only the limited number of
works was devoted to the problem of $1/f^{\alpha}$ noise study and
we have found only a few works where high frequency region of the
tunneling current spectra is studied \cite {Nauen}, \cite {Hohls}.

    We suggest the theoretical
model for tunneling current noise spectra above the flat surface and
above the impurity atoms on semiconductor or metallic surfaces. Our
model gives an opportunity to describe not only singularities in a
low frequency part of tunneling current spectra but also to reveal
singular behavior of tunneling current spectra in a high frequency
region and to describe spectra peculiarities in a wide range of
applied bias voltage. We found experimentally that changing of
tunneling current noise spectra above the flat clean surface and
above the impurity atom depends on the parameters of tunneling
junction such as tip-sample separation or applied bias voltage \cite
{Oreshkin}. Our theoretical model is rather simple, more complicated
models describing noise in semiconductors can be found in \cite
{Galperin}-\cite {Lozano}.

     The investigations of the noise in
two-level system was carried out in \cite {Galperin}. Authors
studied current noise in a double-barrier resonant-tunneling
structure due to dynamic defects that switch states because of their
interaction with a thermal bath. The time fluctuations of the
resonant level result in low-frequency noise, the characteristics of
which depend on the relative strengths of the electron escape rate
and the defect's switching rate. If the number of defects is large,
the noise is of the $1/f^{\alpha}$ type. Shot noise in a mesoscopic
quantum resistor is studied in \cite {Levitov}. Authors found
correlation functions of all order, distribution function of the
transmitted charge and considered Pauli principle as the reason for
the fluctuations. Current fluctuations in a mesoscopic conductor
were analysed in \cite {Altshuler}. Authors derived a general
expression for the fluctuations in the cylindrical tunneling contact
in the presence of a time dependent voltage. In \cite {Moller} noise
of tunneling current at zero bias voltage was investigated. It was
demonstrated that at zero bias voltage the $1/f^{\alpha}$ component
of noise in the tunneling current vanishes and white noise becomes
dominant. The $1/f^{\alpha}$ dependence of the current noise in STM
experiments on graphite in ambient conditions was investigated in
\cite {Tiedje}. Authors attributed this effect to fluctuations
induced by adsorbates in tunneling junction area. In \cite{Lozano}
the fluctuations of tunneling barrier height have been investigated.
The experiments have been performed under UHV conditions on graphite
and gold samples using PtIr tips. From these measurements the
authors have concluded that the intensity of barrier height
fluctuations correspond to the intensity of tunneling current
$1/f^{\alpha}$ noise in the frequency range from 1 to 100 Hz.

  In our previous work we
have found that for the low frequency region  sudden switching on
and off of additional Coulomb potential in tunneling junction area
leads to typical power law dependence of tunneling current noise
spectra at the threshold voltage \cite{Mantsevich}. In this article
our aim is to study one of the possible microscopic origins of
$1/f^{\alpha}$ noise in tunneling contact in the non-resonance cases
and to study high frequency peculiarities of tunneling current
spectra in a wide range of applied bias voltage. We shall analise
modification of tunneling current noise spectrum by the Coulomb
interaction of conduction electrons in the leads (metallic tip and
surface) with non-equilibrium localized charges in tunneling
contact. When electron tunnels from or to localized state the
Coulomb potential is suddenly switched on or off and modifies
tunneling transfer amplitude. It will be shown that corrections to
the tunneling vertexes caused by the Coulomb potential switching on
and off result in nontrivial behavior of tunneling current noise
spectrum in a wide range of applied bias voltage and should be taken
in account. It was theoretically proved and observed experimentally
that the same effects can lead to power law singularity in the
current-voltage characteristics not only the threshold voltage but
also at the high frequency region of tunneling current spectra \cite
{Oreshkin}.
    \section{The suggested model and main results}
 We shall analyse model of two localized states in tunneling contact.
In this case one of the localized states is formed by the impurity
atom in semiconductor and the other one by the tip apex.

    When electron tunnels in or from localized state, the electron
 filling numbers of localized state rapidly change leading to
 appearance of localized state additional charge and sudden
 switching "on" and "off" Coulomb potential. Electrons in the
 leads feel this Coulomb potential.

The model system (Fig.~1) can be described by hamiltonian
$\Hat{H}$:

$$\Hat{H}=\Hat{H}_{0}+\Hat{H}_{tun}+\Hat{H}_{int}$$
\begin{eqnarray}
&\Hat{H}_{0}&=\sum_{p}(\varepsilon_{p}-eV)c_{p}^{+}c_{p}+\sum_{k}\varepsilon_{k}c_{k}^{+}c_{k}+\sum_{i=1,2}\varepsilon_{i}a_{i}^{+}a_{i}\nonumber\\
&\Hat{H}_{tun}&=\sum_{k,i}T_{ki}c_{k}^{+}a_{i}+\sum_{p,i}T_{pi}c_{p}^{+}a_{i}+T\sum
a_{1}^{+}a_{2}+h.c.\nonumber\\
&\Hat{H}_{int}&=\sum_{k,k^{'}}W_{1}c_{k}^{+}c_{k^{'}}a_{1}a_{1}^{+}+
W_{2}c_{k}^{+}c_{k^{'}}a_{2}a_{2}^{+}\nonumber
\end{eqnarray}

 $\Hat{H}_{0}$ describes free electrons in the leads and in the
localized states. $\Hat{H}_{tun}$ describes tunneling transitions
between the leads through localized states. $\Hat{H}_{int}$
corresponds to the processes of intraband scattering caused by
Coulomb potentials $W_{1}$, $W_{2}$ of localized states charges.

 Operators $c_{k}^{+}(c_{k})$ and $c_{p}^{+}(c_{p})$ correspond to
electrons in the leads and operators $a_{i}^{+}(a_{i})$ correspond
to electrons in the localized states with energy
$\varepsilon_{i}$.

\begin{figure}[h]
\centering
\includegraphics[width=70mm]{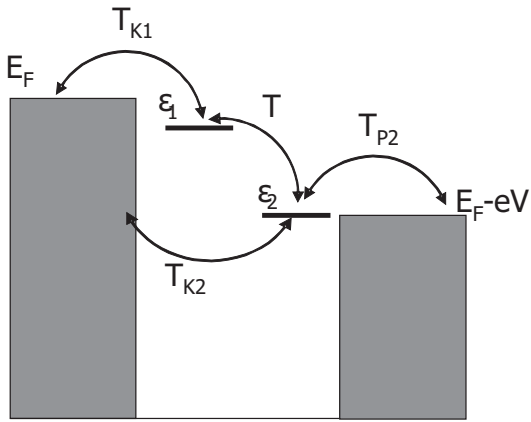}%
\caption{ Schematic diagram of tunneling processes through states
localized on impurity atom and on the STM tip apex.}
\end{figure}

 Current noise correlation function is determined as:
\begin{eqnarray}
(\hbar/e)^{2}\cdot S(t,t')=<I_{L}(t)\cdot{I_{L}(t')}>=\nonumber\\
=\sum_{k,k^{'},i,j}T_{k}^{2}<c_{k}^{+}(t')a_{i}(t')a_{j}^{+}(t)c_{k_{'}}^{+}(t)>
\nonumber\
\end{eqnarray}
where
\begin{eqnarray}
I_{L}(t)=\sum_{k}\dot{n}_{k}\cdot
e=(\sum_{k}c_{k}^{+}(t')a_{i}(t')T_{ki}-h.c.)\cdot\frac{e}{\hbar}\nonumber\
\end{eqnarray}
 The current noise spectra is determined by Fourier
transformation of $S(t,t')$: $S(\omega)=\int S(\tau)d\tau\cdot
e^{i\omega\tau}$. We shall use Keldysh diagram technique in our
study of low frequency tunneling current noise spectra \cite
{Keldysh}.

Let's consider that STM tip is a metal with a cluster on the tip
apex. In this case tip apex localized state energy coincides with
tip Fermi level and consequently $eV=\varepsilon_{2}$.

    When we measure tunneling current spectrum above flat clean surface
we deal with localized state connected with the tip. Tip apex
localized state is usually formed by a cluster of several atoms or
even by a single atom nearest to the surface. Tunneling transfer
amplitude between the sample and the tip decreases exponentially
with increasing of tip-sample separation. Thus in STM junction
effective tunneling from or to the sample occurs through the atomic
cluster nearest to the sample surface.  So the tip apex cluster can
be considered as an intermediate system for electron tunneling from
the sample to the rest of the tip. Tip-sample separation is of order
of interatomic distances. That's why the electron hopping from tip
apex cluster to the bulk of the tip can be also treated as tunneling
process just in the same manner as it is done in tight-binding
approach.

 In our case of weak interaction between localized states
$(T<\gamma_{k1},\gamma_{k2},\gamma_{p2})$ possible variants for
localized states energy levels position in tunneling contact are: 1.
impurity atom localized state energy level exceeds tip apex
localized state energy level ($\varepsilon_{1}>\varepsilon_{2}=eV)$,
2. tip apex localized state energy level  exceeds impurity atom
localized state energy level ($\varepsilon_{2}=eV>\varepsilon_{1}$).

    Expression which describes tunneling current noise spectra without
Coulomb re-normalization can be calculated with the help of Keldysh
diagram technic \cite{Mantsevich}. It consists of three parts.

\begin{figure}[h]
\centering
\includegraphics[width=70mm]{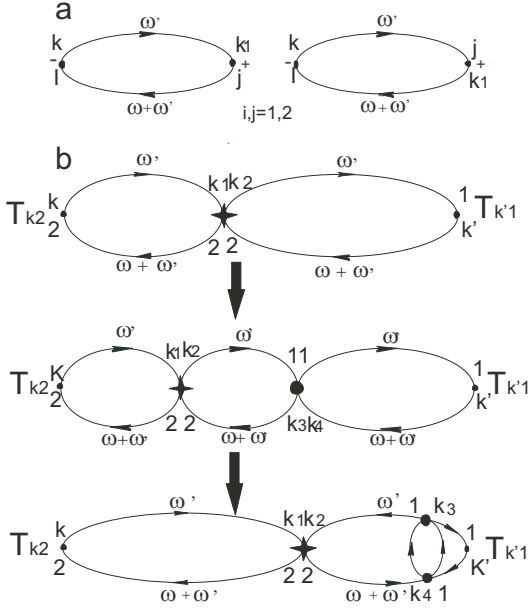}%
\caption{Lowest order diagrams contributing to tunneling current
noise spectra for two localized states in tunneling contact.
 a) In the absence of Coulomb re-normalization of tunneling vertexes.
 b) In the presence of Coulomb re-normalization of tunneling vertexes.
 Tunneling vertexes are marked by the dot.
 Coulomb potential $W_{2}$  is marked by the star. Solid lines correspond to electron Green functions.}
\end{figure}

\begin{eqnarray}
\widetilde{S}_{0}(\omega)=\widetilde{S}_{01}(\omega)+\widetilde{S}_{02}(\omega)+\widetilde{S}_{03}(\omega)\nonumber
\end{eqnarray}

 $\widetilde{S}_{01}(\omega)$ and $\widetilde{S}_{02}(\omega)$ are rather simple
 parts and they have the form:
\begin{eqnarray}
&(\label{one} &\hbar/e)^{2}\cdot
S_{0i}(\omega)=\gamma_{ki}^{2}\cdot\int
d\omega'ImG_{ii}^{R}(\omega')\cdot\nonumber\\
&\cdot& ImG_{ii}^{R}(\omega+\omega')\cdot
(n_{i}(\omega+\omega')-1)\cdot\nonumber\\
&\cdot&(n_{i}(\omega')-n_{k}(\omega'))+
n_{i}(\omega')\cdot(n_{i}(\omega+\omega')-\nonumber\\
&-&n_{k}(\omega+\omega')+\gamma_{ki}^{2}\cdot\int
d\omega'ImG_{ii}^{R}(\omega')\cdot\nonumber\\
&\cdot&ImG_{ii}^{R}(\omega+\omega')\cdot
(n_{k}(\omega+\omega')-1)\cdot n_{i}(\omega')-\nonumber\\
&-&n_{i}(\omega')\cdot(n_{i}(\omega+\omega')-1)-
n_{k}(\omega+\omega')\cdot\
 n_{k}(\omega')+\nonumber\\
&+&n_{k}(\omega')\cdot(n_{i}(\omega+\omega')+ \gamma_{ki}\cdot\int
d\omega'ImG_{ii}^{R}(\omega+\omega')\cdot\nonumber\\
&\cdot&(n_{k}(\omega'))\cdot(n_{i}(\omega+\omega')-1)+
ImG_{ii}^{R}(\omega')\cdot\nonumber\\
&\cdot&(n_{i}(\omega'))\cdot(n_{k}(\omega+\omega')-1)=\widetilde{S}_{0i}\nonumber\
\end{eqnarray}
where $i=1$ corresponds to the  $\widetilde{S}_{01}(\omega)$ and
$i=2$ corresponds to the  $\widetilde{S}_{02}(\omega)$.
 $\widetilde{S}_{03}(\omega)$ is not
trivial, it exists only due to electron tunneling transitions from
one lead to both localized states. Green functions shown on the
graphs are found in \cite {Maslova}. The contribution of
$\widetilde{S}_{01}(\omega)$ is given by graphs with $i=j=1$
(Fig.~2a), $\widetilde{S}_{02}(\omega)$ is described by graph with
$i=j=2$ (Fig.~2a), $\widetilde{S}_{03}(\omega)$ is given by diagrams
with $i\neq j$ (Fig.~2a).

\begin{eqnarray}
&(&\hbar/e)^{2}\cdot
S_{03}(\omega)=8\cdot\gamma_{k_{1}}\cdot\gamma_{k_{2}}\cdot \int
d\omega'ImG_{11}^{R}(\omega')\cdot\nonumber\\
&\cdot&ImG_{22}^{R}(\omega+\omega')\cdot
(n_{1}(\omega')\cdot(n_{2}(\omega+\omega')-1)+\nonumber\\&+&
n_{k}(\omega')\cdot(n_{2}(\omega+\omega')-1)+
(n_{1}(\omega')\cdot\nonumber\\
&\cdot&(n_{k}(\omega+\omega')-1)- n_{k}(\omega')\cdot
(n_{k}(\omega+\omega')-1)+\nonumber\\
&+&8\cdot\gamma_{k_{1}}\cdot\gamma_{k_{2}}\cdot\int
d\omega'ImG_{22}^{R}(\omega')\cdot
ImG_{11}^{R}(\omega+\omega')\cdot\nonumber\\
&\cdot&(n_{2}(\omega')\cdot(n_{1}(\omega+\omega')-1)+\nonumber\\
&+&n_{k}(\omega')\cdot(n_{1}(\omega+\omega')-1)+
(n_{2}(\omega')\cdot\nonumber\\
&\cdot&(n_{k}(\omega+\omega')-1)-n_{k}(\omega')\cdot(n_{k}(\omega+\omega')-1)=\widetilde{S}_{03}\nonumber\
\end{eqnarray}

    Some typical low frequency tunneling current noise spectra for
different values of dimensionless kinetic parameters without Coulomb
re normalization are shown on (Fig.~3a).

   It is clearly evident that when frequency aspire to zero tunneling
current spectra aspire to constant value for different dimensionless
kinetic parameters. When impurity atom energy level exceeds tip apex
localized state energy level ($\varepsilon_{1}>\varepsilon_{2}$) we
have the largest contribution to tunneling current spectra at the
high frequency range of tunneling current spectra
($\omega=\varepsilon_{2}-\varepsilon_{1}$). When tip apex localized
state energy level exceeds impurity atom energy level
($\varepsilon_{2}>\varepsilon_{1}$) we have the largest contribution
to tunneling current spectra at zero frequency.

    Now let us consider re-normalization of the tunneling
amplitude and vertex corrections to the tunneling current spectra
caused by Coulomb interaction between both localized states and
tunneling contact leads. Re-normalization gives us two types of
diagrams contributing to the final tunneling current noise spectra
expression. Ladder diagrams is the most simple type of diagrams
which gives logarithmic corrections to vertexes. But this is not the
only relevant kind of graphs. We must consider one more type of
graphs (parquet graphs) which also gives logarithmicaly large
contribution to tunneling spectra. In parquet graphs a new type of
"bubble" appears instead of "dots" in ladder diagrams. In this
situation one should retain in the n-th order of perturbation
expansion the most divergent terms (Fig.~2).

    It is necessary to re-normalize each vertex individually and
re-normalize both vertexes jointly (Fig.~2b) \cite{Mantsevich}.

    The final expression for tunneling current noise spectra after
Coulomb re-normalization of tunneling vertexes can be written as:

\begin{eqnarray}
&(&\hbar/e)^{2}\cdot
S(\omega)=\widetilde{S}_{0}(\omega)+\widetilde{S}_{01}(\omega)\cdot\nonumber\\
&\cdot&((\frac{D^{2}}{(\omega+eV+E_{1})^{2}+\Gamma_{2}^{2}})^{W_{1}\nu}+\nonumber\\
&+&(\frac{D^{2}}{(\omega+eV+E_{2})^{2}+\Gamma_{1}^{2}})^{W_{1}\nu})+\nonumber\\
&+&\widetilde{S}_{02}(\omega)\cdot
((\frac{D^{2}}{(-\omega+eV+E_{1})^{2}+\Gamma_{1}^{2}})^{W_{2}\nu}+\nonumber\\
&+&(\frac{D^{2}}{(-\omega+eV+E_{2})^{2}+\Gamma_{2}^{2}})^{W_{2}\nu})+
\widetilde{S}_{03}(\omega)\cdot\nonumber\\
&\cdot&((\frac{D^{2}}{(\omega+eV+E_{1})^{2}+\Gamma_{2}^{2}})^{W_{1}\nu}+\nonumber\\
&+&(\frac{D^{2}}{(\omega+eV+E_{2})^{2}+\Gamma_{1}^{2}})^{W_{1}\nu})\cdot\nonumber\\
&\cdot&((\frac{D^{2}}{(-\omega+eV+E_{1})^{2}+\Gamma_{1}^{2}})^{W_{2}\nu}+\nonumber\\
&+&(\frac{D^{2}}{(-\omega+eV+E_{2})^{2}+\Gamma_{2}^{2}})^{W_{2}\nu})\nonumber
\end{eqnarray}

where,

\begin{eqnarray}
E_{1,2}&=&-\frac{\varepsilon_{1}+\varepsilon_{2}}{2}\pm\frac{\sqrt{(\varepsilon_{1}-\varepsilon_{2})^{2}+4T^{2}}}{2}\nonumber\\
\Gamma_{1}&\sim&\Gamma_{2}\sim(\gamma_{k2}+\gamma_{p2}+\gamma_{k1})\nonumber\
\end{eqnarray}

where $D$- are the bandwidths of right and left leads, $\nu$- the
equilibrium density of states in the tunneling contact leads, $W$-
Coulomb potential.
\begin{figure*}
\centering
\includegraphics{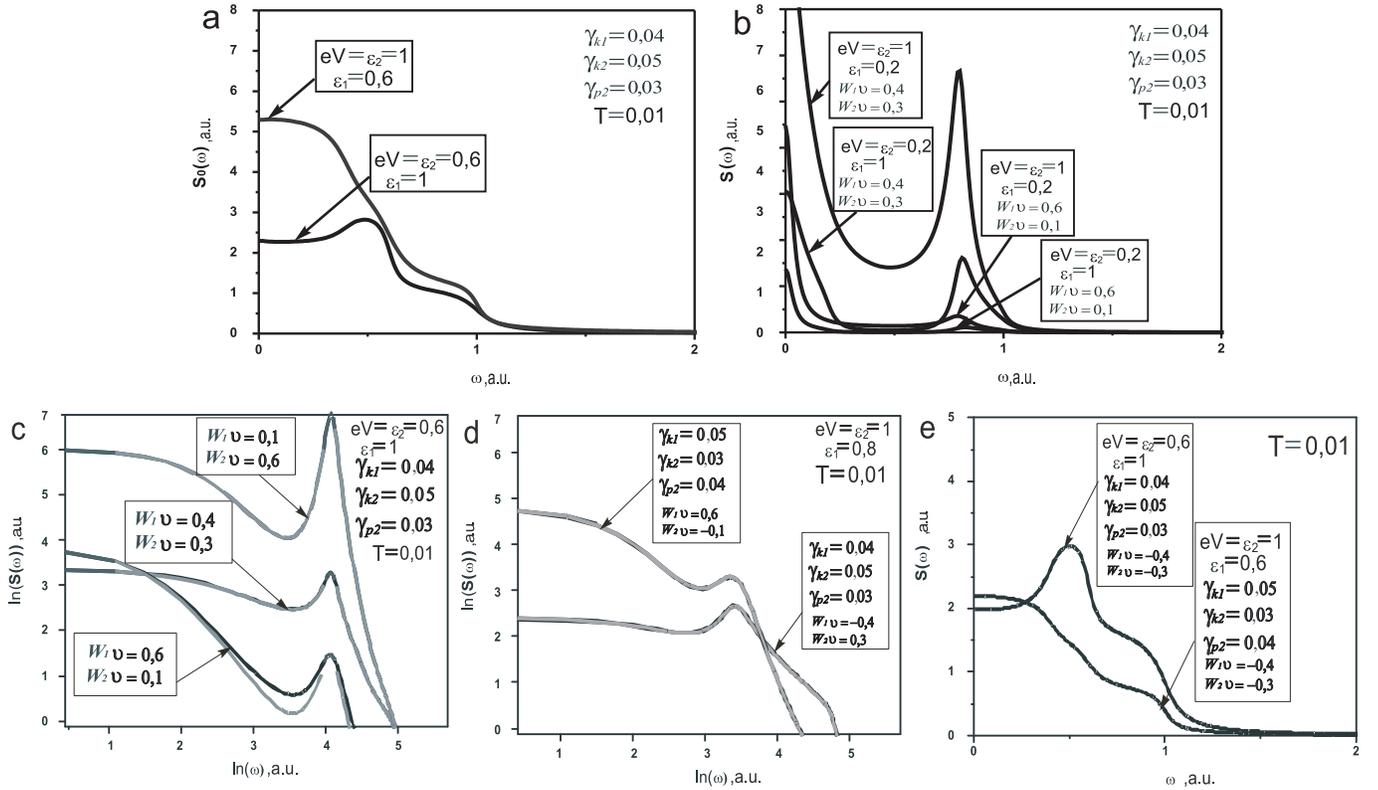}%
\caption{ Typical tunneling current noise spectra for different
 values of dimensionless kinetic parameters for two localized states in tunneling contact ($eV=\varepsilon_{2}\neq\varepsilon_{1}$ ).
 a) In the absence of Coulomb re-normalization of tunneling vertexes.
 b) In the presence of Coulomb re-normalization of tunneling vertexes.
 c) In the presence of
Coulomb re-normalization of tunneling vertexes in double logarithmic
scale when both localized states acquire positive charge.
 d) In the presence of Coulomb re-normalization of tunneling vertexes in double logarithmic scale when localized states acquire charges of different signs.
 e). In the presence of Coulomb re-normalization of tunneling vertexes when both localized states acquire negative charge.}
\end{figure*}

    In the case of strong interaction between localized states
in tunneling contact we have localized states energy levels
splitting and as a result there is no singularity in tunneling
current spectra at the low frequency region. In our situation of
weak interaction between localized states
$(T<\gamma_{k1},\gamma_{k2},\gamma_{p2})$ tunneling transfer
amplitude $T$ plays an important role in kinetic processes in the
tunneling junction but weakly influences on the energy spectra.

   First of all we consider the situation when both localized states
acquire positive charge. Fig.~3b demonstrate low frequency tunneling
current noise spectra for typical values of dimensionless kinetic
parameters. We can see that re-normalization of tunneling matrix
element by switched "on" and "off" Coulomb potential of charged
impurities leads to typical power law singularity in low frequency
part of tunneling current noise spectra and to the peak in the
particular high frequency region, caused by the singularity on the
frequency $\omega=\varepsilon_{2}-\varepsilon_{1}$. Peak amplitude
depends on kinetic parameters and on the charged localized states
Coulomb potentials.

     At the fixed composition of kinetic parameters maximum peak
amplitude corresponds to the case when charged impurity atom
localized state Coulomb potential $W_{1}$ exceeds  Coulomb potential
of tip apex localized state $W_{2}$. In the case of $W_{2}>W_{1}$
(Coulomb potential of charged impurity atom is larger than Coulomb
potential of charged tip apex localized state) peak amplitude
slightly changes by comparison with the case when Coulomb potential
of charged impurity atom is similar to Coulomb potential of charged
tip apex localized state $W_{1}\sim W_{2}$.

    Tunneling current noise spectra in double logarithmic scale demonstrate
 frequency regions where every part of final expression which include
 different power law exponent, approximate noise spectra in the best way Fig.~3c.

    Let's analyse tunneling current spectra shown on Fig.~3c-e.
In the case of two interacting positively charged localized states
the tunneling current spectra at low frequency and in the region of
the high frequency singularity is always determined by the term
which produces the most strong of logarithmic singularity,
determined by the sum of localized states Coulomb potentials at any
values of dimensionless tunneling rates of tunneling contact.
(Fig.~3c).

    If one of the Coulomb potentials strongly exceeds the other one
 this potential determine tunneling current noise spectra with increasing of frequency.
If Coulomb potential of charged impurity atom is similar to Coulomb
potential of charged tip apex localized state $W_{1}\sim W_{2}$,
tunneling current noise spectra with increasing of frequency is
determined by Coulomb potential of charged tip apex state $W_{2}$
(Fig.~3c).

    Let us consider tunneling current noise spectra in double logarithmic scale in the
situation when localized states acquire charges of different signs
Fig.~3d.

    Tunneling current noise spectra in double logarithmic scale
(Fig.~3d) make it clear that when localized states acquire charges
of opposite signs tunneling current noise spectra in the low
frequency region and in the region of the high frequency singularity
($\omega=\varepsilon_{2}-\varepsilon_{1}$)  are always approximated
by the term depending on maximum positive value of impurity atom
Coulomb potential $W_{1}$ or tip apex localized state Coulomb
potential $W_{2}$.
    Tunneling current noise spectra in double logarithmic scale in the situation when both
localized states acquire negative charges (Fig.~3e) make it clearly
evident that singular effects become negligible.

    Now let's describe interaction effects in tunneling contact when
both localized states are formed by impurity atoms in the surface.
In this case we can say about cluster in the surface
$eV\neq\varepsilon_{2}\neq\varepsilon_{1}$.

    Possible variants for localized states energy levels position in tunneling contact are:

1. applied bias voltage exceeds energy levels of both localized
states in tunneling contact ($eV>\varepsilon_{2}>\varepsilon_{1}$,
$eV>\varepsilon_{1}>\varepsilon_{2}$); 2. applied bias voltage
exceeds energy level of one of the localized states in tunneling
contact ($\varepsilon_{2}>eV>\varepsilon_{1}$,
$\varepsilon_{1}>eV>\varepsilon_{2}$); 3. energy levels of both
localized states in tunneling contact exceed applied bias voltage
($\varepsilon_{2}>\varepsilon_{1}>eV$,
$\varepsilon_{1}>\varepsilon_{2}>eV$)

    Typical low frequency tunneling current noise spectra for
different values of dimensionless kinetic parameters without Coulomb
re-normalization are shown on Fig.~4a.
\begin{figure*}
\centering
\includegraphics{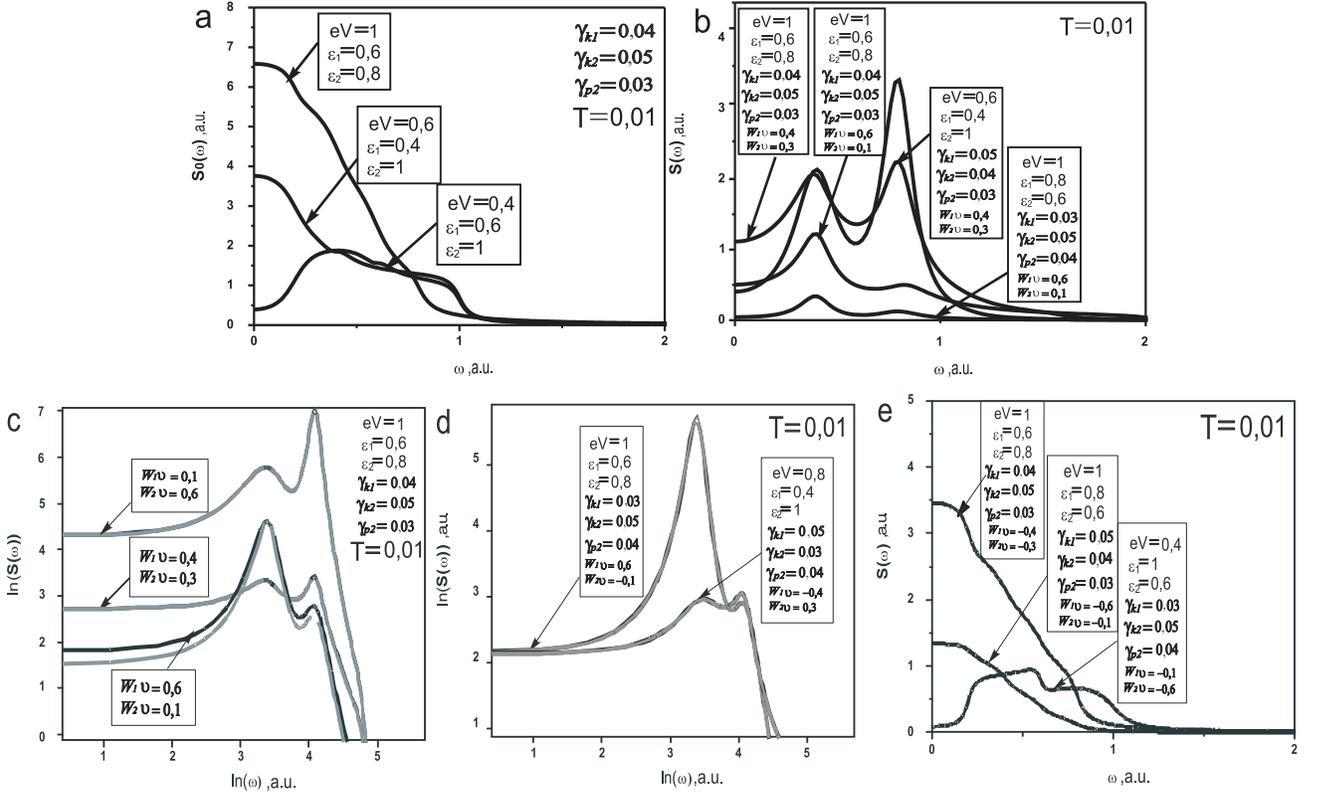}%
\caption{ Typical tunneling current noise spectra for different
values of dimensionless kinetic parameters for two localized states
in tunneling contact ($eV\neq\varepsilon_{2}\neq\varepsilon_{1}$).
a) In the absence of Coulomb re-normalization of tunneling vertexes.
b) In the presence of Coulomb re-normalization of tunneling
vertexes. c) In the presence of Coulomb re-normalization of
tunneling vertexes in double logarithmic scale when both localized
states acquire positive charge. d)In the presence of Coulomb
re-normalization of tunneling vertexes in double logarithmic scale
when localized states acquire charges of different signs. e)In the
presence of Coulomb re-normalization of tunneling vertexes in double
logarithmic scale when both localized states acquire negative
charge. }
\end{figure*}

    It is clearly evident that when frequency aspire to zero tunneling
current spectra aspire to constant value for different dimensionless
kinetic parameters.
    When energy levels of both localized
states in tunneling contact exceeds applied bias voltage
($\varepsilon_{2}>\varepsilon_{1}>eV$ or
$\varepsilon_{1}>\varepsilon_{2}>eV$) we have the largest
contribution to tunneling current spectra in the high frequency
region (Fig.~4a). In the other cases we have the largest
contribution to tunneling current spectra at zero frequency
(Fig.~4a).

    Now let us consider re-normalization of the tunneling amplitude and vertex corrections
 to the tunneling current spectra caused by Coulomb interaction between both localized states and tunneling contact leads.
Fig.~4b demonstrate tunneling current noise spectra for typical
values of dimensionless kinetic parameters. It is clearly evident
that when frequency aspire to zero tunneling current spectra aspire
to constant value and we have no power law singularity in a low
frequency part of tunneling current spectra.

    We can see that re-normalization of tunneling matrix element by
switched "on" and "off" Coulomb potential of charged impurities
leads to the singular peaks in high frequency regions of tunneling
current spectra, caused by singularities on the frequencies
$\omega=eV-\varepsilon_{1}$ and $\omega=eV-\varepsilon_{2}$.

    Let's start from the situation when both localized states acquire positive charge.
Tunneling current noise spectra in double logarithmic scale
demonstrate frequency regions where every part of final expression
which include different power law exponent, approximate noise
spectra in the best way Fig.~4c.

    Let's analyze tunneling current spectra shown on Fig.~4c-e. In the
case of two interacting positively charged localized states the
tunneling current spectra in the regions of the singular peaks
 is determined by
the term which produces the most strong of logarithmic singularity,
determined by the sum of localized states Coulomb potentials at any
values of dimensionless tunneling rates of tunneling contact.
(Fig.~4c).

    If one of the Coulomb potentials strongly exceeds the other one this
potential determine tunneling current noise spectra except frequency
regions in the vicinity of singular peaks . If Coulomb potentials
have the similar values $W_{1}\sim W_{2}$, tunneling current noise
spectra is determined by Coulomb potential of charged tip apex state
$W_{2}$ (Fig.~4c).

    Let us consider tunneling current noise spectra in double
logarithmic scale in the situation when localized states acquire
charges of different signs (Fig.~4d).

    Tunneling current noise spectra in double logarithmic scale
(Fig.~4d) make it clear that when localized states acquire charges
of opposite signs tunneling current noise spectra is always
approximated by the term depending on maximum positive value of
impurity atom Coulomb potential $W_{1}$ or tip apex localized state
Coulomb potential $W_{2}$.

    Tunneling current noise spectra in double logarithmic scale in the
situation when both localized states acquire negative charges
(Fig.~4e) make it clearly evident that singular effects after
Coulomb re-normalization become negligible. Simple estimation gives
possibility to analyze the validity of obtained results. For typical
composition of tunneling junction parameters in the low frequency
region power spectrum of tunneling current corresponds to
experimental results. Power spectrum on zero frequency has the form:
\begin{eqnarray}
S(0)\approx(\gamma_{eff}\cdot
e/\hbar)^{2}\cdot(D/\gamma_{eff1})^{\nu\cdot W}\cdot
(1/\bigtriangleup\omega)\nonumber
\end{eqnarray}
For typical $\gamma_{eff}$, $\gamma_{eff1}$ $\approx10^{-13}$,
$D\approx10$, $W\approx0,5$ $S(0)\approx10^{-18} A^{2}/Hz$ \cite
{Oreshkin}.

\section{Conclusion}
  The microscopic theoretical approach describing tunneling current
noise spectra in a wide range of applied bias voltage taking in
account many-particle interaction was proposed. When electron
tunnels to or from localized state the charge of localized state
rapidly changes. This results in sudden switching on and off of
additional Coulomb potential in tunneling junction area, and leads
to singular behavior of tunneling current spectra in a particular
frequency range determined by the parameters of tunneling junction
such as applied bias voltage or the localized states energy levels
deposition.

    In the case of two localized states in tunneling junction when energy level of one of
the localized states is connected with the tip apex and is not equal
to energy level of the impurity atom localized state on the surface
we can see typical power law dependence for the low frequency part
of tunneling current spectra and singular peak  in the particular
high frequency region of tunneling current spectra determined by the
localized states energy levels deposition in the tunneling junction
($eV=\varepsilon_{2}\neq\varepsilon_{1}$). In the non-resonance case
of the cluster on the surface
($eV\neq\varepsilon_{2}\neq\varepsilon_{1}$) we have found two
singular peaks in the high frequency region of tunneling current
noise spectra. So our results demonstrate that changing of the
applied bias voltage leads to the tuning of tunneling current noise
spectra .

    This effect can be
qualitatively understood by the following way: apart from the direct
tunneling from the localized state to the state with momentum $k$ in
the lead of tunneling contact described by the amplitude $T_{k}$,
the electron can first tunnel into any other empty state $k'$ in the
lead and then scatter to the state k by Coulomb potential. So the
appearance of even a weak Coulomb potential significantly modifies
the many particle wave function of the electron gas in the leads.

 We are grateful to A.I. Oreshkin and S.V. Savinov for discussions
 and useful remarks.

This work was partially supported by RFBR grants ¹ 06-02-17076-a, ¹
06-02-17179-a, ¹ 08-02-01020-a and the Council of the President of
the Russian Federation for Support of Young Scientists and Leading
Scientific School ¹ NSh-4464.2006.2.

\pagebreak

\end{document}